\documentclass[prd, preprint, nofootinbib, showpacs]{revtex4}

\usepackage{epsf,epsfig,subfigure,graphicx,amsmath,amssymb}
\usepackage{amsfonts}
\usepackage{latexsym}
\usepackage{color}

\newcommand{\dis}[1]{\begin{equation}\begin{split}#1\end{split}\end{equation}}
\newcommand{\be}{\begin{equation}}
\newcommand{\ee}{\end{equation}}
\def\bea{\begin{eqnarray}}
\def\eea{\end{eqnarray}}

\newcommand{\eq}[1]{Eq.~(\ref{#1})}

\newcommand{\bfrac}[2]{{\left(\frac{#1}{#2} \right)  }}\newcommand{\VEV}[1]{\langle #1 \rangle}

\newcommand\gev{\,{\rm GeV}}

\begin{document}

\title{
 Dirac right-handed sneutrino dark matter and its signature in the gamma-ray lines
}

\author{Ki-Young Choi}
 \affiliation{
 Asia Pacific Center for Theoretical Physics, Pohang, Gyeongbuk 790-784, Republic of Korea and \\
 Department of Physics, POSTECH, Pohang, Gyeongbuk 790-784, Republic of Korea}

\author{Osamu Seto}
 \affiliation{
 Department of Life Science and Technology,
 Hokkai-Gakuen University,
 Sapporo 062-8605, Japan
}

%

\begin{abstract}
%
We show that a Dirac right-handed scalar neutrino can be dark matter (DM) as a weakly interacting massive particle
 in the neutrinophilic Higgs model. 
When the additional Higgs fields couple only to the leptonic sector through neutrino Yukawa couplings,
 the right number of relic density of dark matter can be obtained from thermal freeze-out of
 the dark matter annihilation into charged leptons and neutrinos.
At present epoch, this tree-level annihilation into fermions is suppressed by the velocity of dark matter,
  and the one-loop annihilation cross section into $\gamma\gamma$ can be dominant
  because relevant coupling constants are different.
Hence, the recently observed (tentative) gamma-ray line signal in the Fermi-Fermi-Large Area Telescope can be naturally explained
by the annihilation of right-handed sneutrino dark matter.
%
\end{abstract}


\preprint{APCTP-Pre2012-008, HGU-CAP-017} 

\vspace*{3cm}
\maketitle


\section{Introduction}
\label{introduction}

Various cosmological and astrophysical observations show convincing evidence of 
 nonbaryonic dark matter (DM).
However, the identity of the dark matter still remains one of the 
 most signiﬁcant unanswered questions
in particle physics and astronomy. 
One of the most promising and natural candidates may be the 
weakly interacting massive particles (WIMPs) in the physics beyond the standard model~\cite{DMreview}.

Much effort has been made in experiments for the direct and indirect detection of WIMP.
The strategy for the direct detection of dark matter is to look for the recoil energy from scattering
 off with nuclei by WIMPs,
 while that of indirect detection is to find an excess over the  astrophysical background of cosmic rays
 due to additional products by dark matter annihilation or decay.
Those annihilation products include neutrinos, positrons, antiprotons, gamma rays and so on. 

In general, the gamma-ray flux from dark matter  annihilation or decay consists of two components: 
 the continuous spectrum and line spectrum.
A gamma-ray line is a clear signature of the WIMP annihilation and
 has been regarded as a ``smoking gun'' for the WIMP dark matter 
 because there is no known astrophysical source that can emit such a line gamma ray.
However, a tentative indication of gamma-ray line using Fermi-LAT data was reported recently
 in~\cite{Bringmann:2012vr,Weniger:2012tx}. 
It can be associated with dark matter
 annihilation~\cite{Bringmann:2012vr,Weniger:2012tx,Ibarra:2012dw,Dudas,Cline}
 or due to hard photons in the Fermi bubbles regions~\cite{Su:2010qj,Profumo:2012tr}: 
 however, Ref.~\cite{Tempel} confirms the existence of spectral feature
 and finds no correlation with Fermi bubbles.

In general, the WIMP annihilation cross section into gamma-ray line is small.
Since a dark matter candidate by definition does not couple to a photon directly,
 those annihilation modes are induced by loop processes, and hence, suppressed
 compared to that of tree level. 
 There are a few nonsupersymmetric models which produce gamma-ray line,
 e.g., the singlet scalar DM~\cite{Profumo:2010kp},
 the loop induced neutrino mass model~\cite{Aoki:2009vf}, and
 the inert Higgs doublet model~\cite{Gustafsson:2007pc}.
However, it is not easy to produce in the supersymmetric models. 
In fact, the expected gamma-ray line flux, in particular $\gamma\gamma$, induced
 by the annihilation of neutralino in the minimal supersymmetric standard model (MSSM)
 is much suppressed~\cite{BHS}.
Another candidate of WIMP dark matter in the supersymmetric models can be sneutrino: 
 mixed~\cite{ArkaniHamed:2000bq} or right-handed sneutrino~\cite{Lee:2007mt,Cerdeno:2008ep,Bandyopadhyay:2011qm}.
However, in those models the relevant couplings of  sneutrinos are connected to the neutral Higgs bosons or $Z'$ boson, 
and therefore it is not easy to produce a large line gamma-ray flux from their annihilation
 processes.~\footnote{For the continuous spectrum,
 a possible significant Breit-Wigner enhancement in s-channel processes
 has been pointed out~\cite{CHPS}.}

In this paper, we show that a Dirac right-handed sneutrino as WIMP dark matter could produce
 gamma-ray line flux which is significantly larger than
 in other supersymmetric models.   
Right-handed Dirac sneutrino dark matter has been proposed
 to explain dark matter  in the MSSM by introducing  right-handed neutrino superfields~\cite{Asaka}
 and in a further extended model~\cite{Demir:2009kc},
 but those are not thermal freeze out relics.
The essential ingredient for our Dirac sneutrino to be a WIMP is
 the extended Higgs sector including a neutrinophilic Higgs field, which
 is a Higgs field interacting with other matters only through neutrino Yukawa
 couplings~\cite{Ma,Wang,Ma:2006km,Nandi}.
The neutrinophilic Higgs model
 is based on the concept that the smallness of neutrino mass might
 not come from a small Yukawa coupling
 but a small vacuum expectation value (VEV) of the neutrinophilic Higgs field $H_{\nu}$.
Recently, various aspects of neutrinophilic Higgs models have been
 studied~\cite{Davidson:2009ha,Logan:2010ag,HabaHirotsu,Sher:2011mx,HabaSeto,
Haba:2011nb,Haba:2011fn,Haba:2012ai}.
The consequence of the neutrinophilic Higgs model is
 that neutrino Yukawa couplings are not necessarily small
 because the smallness of neutrino masses is explained by the small $H_{\nu}$ VEV. 
Actually, neutrino Yukawa couplings can be as large as of the order of unity.
Hence we will show that by using this advantage,
 right-handed Dirac sneutrino can have a large enough annihilation cross section 
 to be WIMP, as well as to produce an observable line gamma-ray flux.

The paper is organized as follows.
In Sec.~\ref{model},
 we briefly describe a supersymmetric neutrinophilic Higgs model where
 the VEV of  $H_{\nu}$ is small and neutrino Yukawa coupling can be large.
In Sec.~\ref{sneutrinoDM}, we examine the Dirac right-handed sneutrino dark matter candidate by
 estimating its thermal relic density and show how
 large monochromatic gamma-ray line signal can be produced.
We then summarize our results in Sec.~\ref{conclusion}.

\section{Model}
\label{model}

The supersymmetric neutrinophilic Higgs model has 
 a pair of neutrinophilic Higgs doublets
 $H_{\nu}$ and $H_{\nu'}$ in addition to 
 up- and down-type two Higgs doublets $H_u$ and $H_d$ in
 the MSSM~\cite{HabaSeto}.
A discrete $Z_2$ parity  is also introduced to discriminate
 $H_u (H_d)$ from $H_{\nu}(H_{\nu'})$,
 and the corresponding charges are assigned in Table~\ref{Table}. 
\begin{table}[t]
\centering
\begin{center}
\begin{tabular}{|c|c|c|} \hline
Fields  &  $Z_{2}$ parity & Lepton number \\ \hline\hline
MSSM Higgs doublets, $H_u, H_d$  &  $+$ &  0 \\ \hline
New Higgs doublets, $H_{\nu}, H_{\nu'}$ 
 &  $-$ & 0 \\ \hline
Right-handed neutrinos, $N$  &  $-$ & $1$ \\ \hline
\end{tabular}
\end{center}
\caption{The assignment of $Z_2$ parity and lepton number.}
\label{Table}
\end{table}
%
Under this discrete symmetry, 
 the superpotential is given by
\begin{eqnarray}
 W &=&y_{u} Q \cdot H_u U_{R}
 +y_d Q \cdot {H_d}D_{R}+ y_l L \cdot H_d E_{R} +y_{\nu} L \cdot H_{\nu} N  \nonumber \\
&& +\mu H_u \cdot H_d + \mu' H_\nu \cdot  H_{\nu'} 
+\rho H_u \cdot H_{\nu'} + \rho' H_\nu \cdot H_d ,
\label{superpotential}
\end{eqnarray}
 where we omit generation indexes and dot represents $SU(2)$ antisymmetric product. 
The $Z_2$ parity plays a crucial role of 
 suppressing tree-level flavor changing neutral currents (FCNCs) and  
 is assumed to be softly broken 
 by tiny parameters of $\rho$ and $\rho' (\ll \mu, \mu' )$.

The scalar potential relevant for Higgs fields and sneutrinos is given by the supersymmetry (SUSY) potential
and SUSY breaking terms,
\begin{eqnarray}
 V &=& V_{\rm SUSY} + V_{\rm soft} ,
 \label{potential}
 \end{eqnarray}
 with
\begin{eqnarray}
 V_{\rm SUSY} &=& |y_l H_d \tilde{E}_R +y_{\nu} H_{\nu} \tilde{N} |^2
    + | y_{\nu} \tilde{L}  \tilde{N} - \mu'  H_{\nu'} +\rho' H_d |^2 +| y_{\nu} \tilde{L} \cdot H_{\nu} |^2
    + | y_l \tilde{L} \cdot H_d |^2  \nonumber \\
  &&  + |\mu H_d +\rho H_{\nu'}|^2 + |y_l \tilde{L} \tilde{E}_R + \mu H_u + \rho' H_\nu |^2
        + | \mu' H_{\nu} +\rho H_u |^2 + {\rm D-terms}, \label{Vsusy}
\end{eqnarray}
and
\begin{eqnarray}
 V_{\rm soft} &=& m_{H_u}^2|H_u|^2 + m_{H_d}^2|H_d|^2
     + m_{H_{\nu}}^2|H_{\nu}|^2 + m_{H_{\nu'}}^2|H_{\nu'}|^2
    + m_{\tilde{L}}^2|\tilde{L}|^2+ m_{\tilde{N}}^2|\tilde{N}|^2  \nonumber \\
&& + ( y_l A_l \tilde{L} \cdot H_d \tilde{E}_R +y_{\nu} A_{\nu} \tilde{L} \cdot H_{\nu} \tilde{N}  \nonumber \\
&& + B \mu H_u \cdot H_d + B' \mu' H_\nu \cdot  H_{\nu'} 
+ B_{\rho} \rho H_u \cdot H_{\nu'} + B_{\rho'} \rho' H_\nu \cdot H_d + {\it  h.c.} ) .\label{Vsof}
\end{eqnarray}
The Higgs dependent part of the scalar potential is expressed as
\begin{eqnarray}
 V &=& | - \mu'  H_{\nu'} +\rho' H_d |^2 + |\mu H_d +\rho H_{\nu'}|^2 + |\mu H_u + \rho' H_\nu |^2
        + | \mu' H_{\nu} +\rho H_u |^2 + {\rm D-terms} \nonumber \\
 && + m_{H_u}^2|H_u|^2 + m_{H_d}^2|H_d|^2
     + m_{H_{\nu}}^2|H_{\nu}|^2 + m_{H_{\nu'}}^2|H_{\nu'}|^2  \nonumber \\
 && + (  B \mu H_u \cdot H_d + B' \mu' H_\nu \cdot  H_{\nu'} 
    + B_{\rho} \rho H_u \cdot H_{\nu'} + B_{\rho'} \rho' H_\nu \cdot H_d + {\it  h.c.} ) .
\end{eqnarray}
The tiny soft $Z_2$-breaking parameters $\rho, \rho' $ generate  
 a large hierarchy of $v_{u,d} (\equiv \langle H_{u,d}\rangle) \gg 
 v_{\nu, \nu'}(\equiv \langle H_{\nu, \nu'}\rangle)$, which are given by the
 stationary conditions 
\begin{eqnarray}
&&
\left(
\begin{array}{cc}
 m_{H_{\nu}}^2 + \mu'^2 + \frac{m_Z^2 }{2}\frac{\tan^2\beta - 1 }{\tan^2\beta + 1 }   & -B_{\mu'} \mu' \\
 -B_{\mu'} \mu' &  m_{H_{\nu'}}^2 + \mu'^2 - \frac{m_Z^2}{2}\frac{\tan^2\beta - 1 }{\tan^2\beta + 1 } 
\end{array}
\right) 
\left(
\begin{array}{c}
 v_{\nu}  \\
 v_{\nu'}
\end{array}
\right) \nonumber \\
 && \simeq 
\left(
\begin{array}{cc}
 - (\mu' \rho + \mu \rho')    & B_{\rho'} \rho' \\
 B_{\rho} \rho &  - (\mu \rho + \mu' \rho') 
\end{array}
\right) 
\left(
\begin{array}{c}
 v_u  \\
 v_d
\end{array}
\right) . 
\end{eqnarray}
Namely, we obtain
\begin{equation}
 v_{\nu} = {\cal O}\left(\frac{\rho}{\mu'}\right) v.
\end{equation}
The hierarchy of $\rho/\mu' \ll 1$ leads to a small $v_{\nu}$ and 
 the smallness of $\rho$ compared to $\mu'$ is explained naturally in 't Hooft's sense 
 because $\rho$ is a soft breaking parameter of the $Z_2$ parity.
It is easy to see that neutrino Yukawa couplings $y_{\nu}$ can be large
 for small $v_{\nu}$ using the relation of the Dirac neutrino mass $m_{\nu} = y_{\nu} v_{\nu}$.
For $v_{\nu} \sim 0.1 $ eV, it gives $y_{\nu} \sim 1$.
 
At the vacuum of $v_{\nu, \nu'}\ll v_{u,d}$ that we are interested in,
 physical Higgs bosons originated from  $H_{u, d}$
 are almost decoupled from
 those from $H_{\nu,\nu'},$ except a tiny mixing
 of the order of ${\cal O}\left(\rho/M_{\rm SUSY}, \rho'/M_{\rm SUSY} \right)$
 where $M_{\rm SUSY} ( \sim 1 $ TeV) denotes the scale of soft SUSY breaking parameters.
The former $H_{u,d}$ doublets almost constitute Higgs bosons in the MSSM -
 two CP-even Higgs bosons $h$ and $H$, 
 one CP-odd Higgs boson $A$, and a charged Higgs boson $H^\pm$ - 
 while the latter, $H_{\nu, \nu'}$, constitutes 
 two CP-even Higgs bosons $H_{2,3}$,
 two CP-odd bosons $A_{2,3}$,
 and two charged Higgs bosons $H^\pm_{2,3}$. 

The scalar potential including sneutrinos is given by
\begin{eqnarray}
 V &=& |y_{\nu} H_{\nu} \tilde{N} |^2
    + | y_{\nu} \tilde{L} \tilde{N} - \mu' H_{\nu'} +\rho' H_d |^2 +| y_{\nu} \tilde{L} \cdot H_{\nu} |^2
    + | y_l \tilde{L} \cdot H_d |^2   \nonumber  \\
   && + m_{L}^2|\tilde{L}|^2 + m_{\tilde{N}}^2|\tilde{N}|^2 
    + ( y_l A_l \tilde{L} \cdot H_d \tilde{E}_R +y_{\nu} A_{\nu} \tilde{L} \cdot H_{\nu} \tilde{N}  + {\it  h.c.} )
    + {\rm D-terms}.
\end{eqnarray}
The mixing between LH and RH sneutrino is roughly estimated as
\begin{equation}
\sin \theta_{\tilde{\nu}} = {\cal O}\left( \frac{m_\nu}{M_{\rm SUSY}} \right) .
\label{LRmixing}
\end{equation}
We find that the RH sneutrino $\tilde{N}$, which is a dark matter candidate in our model, has suppressed interactions
to the SM-like Higgs boson or Z boson  since they are proportional to the mixing of LH and RH neutrinos,
$\sin \theta_{\tilde{\nu}} $ in \eq{LRmixing}.

\section{Right-handed scalar neutrino as dark matter}
\label{sneutrinoDM}

In this section we will investigate the thermal relic abundance of the right-handed sneutrino dark matter ${\tilde N}$
and its indirect signature in the gamma-ray observation.

\subsection{Thermal relic density of dark matter : Tree-level annihilation of ${\tilde N}$}

\begin{figure}[!t]
  \begin{center}
  \begin{tabular}{c}
   \includegraphics[width=0.5\textwidth]{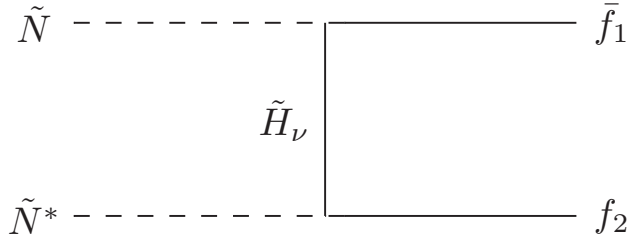}
     \end{tabular}
  \end{center}
  \caption{Tree-level diagram for the annihilation of right-handed sneutrinos.}
  \label{fig:Tree}
\end{figure}

Since the dark matter particle $\tilde N$ can have large Yukawa couplings given
 by $y_\nu\sim {\mathcal O}(1)$, 
 the DMs can be in the thermal equilibrium in the early Universe through
 these large Yukawa interactions.
Here, we consider the case that  the mass eigenstates $H_2$ and $H_3$ originated mostly
 from $H_{\nu}$ and $H_{\nu'}$ are much
 heavier than $M_{\tilde N}$
 then the electroweak precision measurement constraints are easily satisfied 
 and the annihilation into $H_{\nu}$ and $H_{\nu'}$ is kinematically forbidden.
In this case, the dominant annihilation mode of $\tilde{N}$ in the early Universe 
is the annihilation into a lepton pair
 $\tilde{N} \tilde{N}^*\rightarrow \bar{f_1} f_2 $ mediated by the heavy $H_{\nu}$-like
 Higgsinos as described in Fig.~\ref{fig:Tree}. 
 The final states $f_1$ and $f_2$ are charged leptons for the $t$-channel $\tilde{H}_{\nu}$-like
 charged Higgsino ($\tilde{H}_{\nu}^\pm$) exchange, while
those are neutrinos for the $t$-channel $\tilde{H}_{\nu}$-like
 neutral Higgsino ($\tilde{H}_{\nu}^0$) exchange. 

The thermal averaged annihilation cross section for this mode
 in the early Universe is expressed in partial wave expansion method as~\cite{Lindner:2010rr}
\begin{equation}
\langle \sigma v\rangle = \sum_f \left(
 \frac{y_\nu^4}{16\pi} \frac{m_f^2}{(M^2_{\tilde N} + M^2_{\tilde{H}_\nu} )^2}
 + \frac{ y_{\nu}^4}{8 \pi }
    \frac{M^2_{\tilde{N}}}{ ( M_{\tilde{N}}^2 + M_{\tilde{H_{\nu}}}^2 )^2 }\frac{T}{ M_{\tilde{N}} } + ... \right) ,
    \label{treelevel}
\end{equation}
 where we used $\VEV{v_{\rm rel}^2}=6T/M_{DM}$
 with $v_{\rm rel}$ being the relative velocity of annihilating dark matter particles,
 and $m_f$ is the mass of
 the fermion $f$ and $M_{{\tilde H}_\nu} \simeq \mu'$
 denotes the mass of $\tilde{H}_{\nu}$-like Higgsino.
For simplicity we have assumed that Yukawa couplings are universal for each flavor.

\begin{figure}[!t]
  \begin{center}
  \begin{tabular}{c}
   \includegraphics[width=0.6\textwidth]{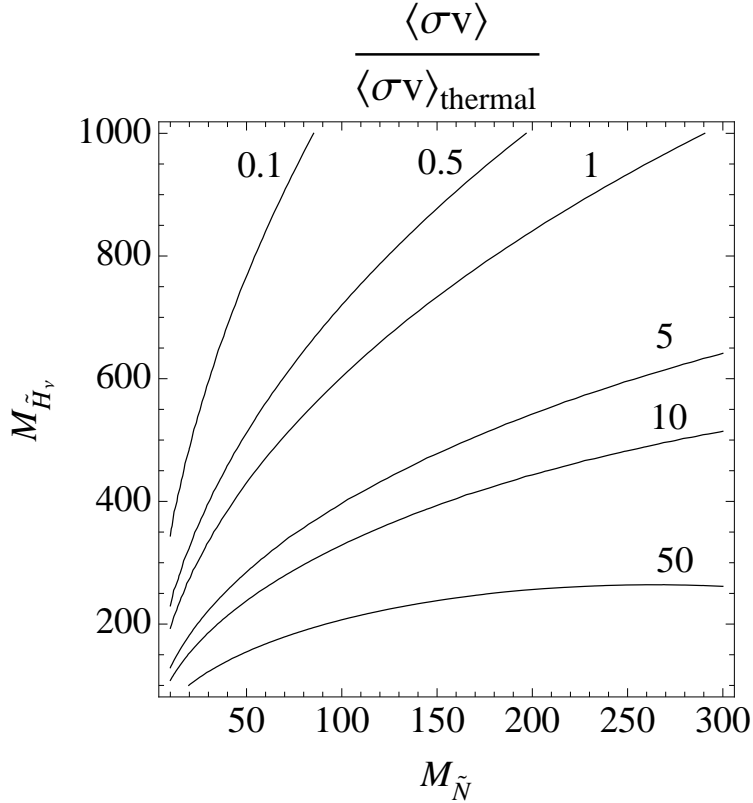}  \end{tabular}
  \end{center}
  \caption{ The thermal averaged annihilation cross section of $\tilde{N}$ at freeze-out temperature normalized by that required for the right relic density of WIMP. Here we used $T/M_{\tilde{N}}=1/20$ and $y_\nu=1$. }
  \label{fig:sigmav}
\end{figure}

Since the $s$-wave contribution of the first term in the right-hand side is helicity suppressed,
 the $p$-wave annihilation cross section of the second term is relevant for the dark matter relic density
 at freeze-out epoch, $T_f\sim M_{\tilde{N}}/20 $.   
The right relic density of WIMP
 can be obtained for the thermal averaged annihilation cross section:
\begin{equation}
\VEV{\sigma v}_{thermal} \simeq 3\times 10^{-26} \, cm^3 \, s^{-1}=2.57\times 10^{-9} \gev^{-2}.
\label{freezeout}
\end{equation}
Comparing Eqs.~(\ref{treelevel}) and (\ref{freezeout}),  we find $y_{\nu} = {\cal O}(1)$ as
\dis{
y_\nu^4\simeq 1.09 \,  \bfrac{130\gev}{M_{\tilde N}}^2 \bfrac{{M}_{\tilde{H}_\nu}}{700\gev}^4\bfrac{\VEV{\sigma v}_{thermal}}{2.57\times 10^{-9}\gev^{-2}} \left(\frac{1/20}{T_f/M_{\tilde N}}\right),
} 
 where $T_f$ is the DM freeze-out temperature, which is usually $T_f\simeq M_{DM}/20$,
 and we accounted for the number of modes of the final states $\sum_f=2\times 3^2=18$.
In Fig.~\ref{fig:sigmav} we show the contour plot of the annihilation cross section in the plane of
 $M_{\tilde{N}}$ and $M_{{\tilde H}_\nu}$ for $y_\nu=1$.

\subsection{Monochromatic gamma-ray lines from right-handed sneutrino annihilation}

Since we are considering the massive dark matter, which is nonrelativistic at present, the tree-level p-wave contribution to the annihilation of DM in \eq{treelevel} is also suppressed. 
Therefore, in this model, the dominant contribution to the annihilation of DM in the galaxy at present universe comes from the loop diagrams. 

The emission of a vector boson through the virtual internal bremsstrahlung can
 enhance the $s$-wave contribution in particular when the mass splitting between dark matter  and the $t$-channel mediator, $H_\nu$
 in our case, is small. 
However since we are considering  heavy Higgs, $M_{H_\nu}\gg M_{\tilde N}$, the bremsstrahlung is suppressed. 
Therefore we do not expect the line spectrum of gamma ray from internal bremsstrahlung, 
 which is different from that in~\cite{Bringmann:2012vr}.

\begin{figure}[!t]
  \begin{center}
  \begin{tabular}{ccc c ccc}
   \includegraphics[width=0.3\textwidth]{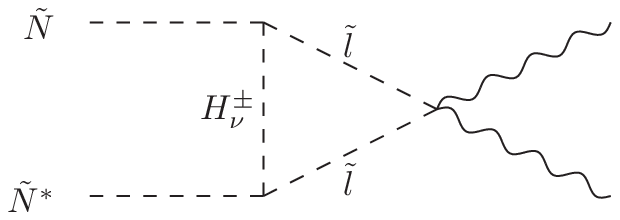}
    & & & &&&
  \includegraphics[width=0.3\textwidth]{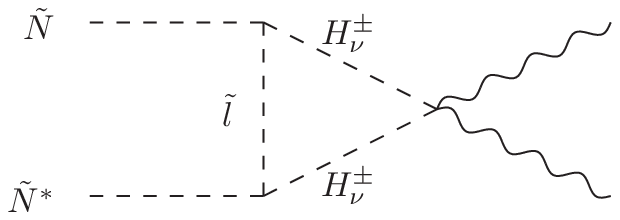}
  \end{tabular}
  \end{center}
  \caption{Triangle loop diagrams for the annihilation of right-handed sneutrinos.}
  \label{fig:Triangle}
\end{figure}
\begin{figure}[!t]
  \begin{center}
  \begin{tabular}{ccc c ccc}
   \includegraphics[width=0.3\textwidth]{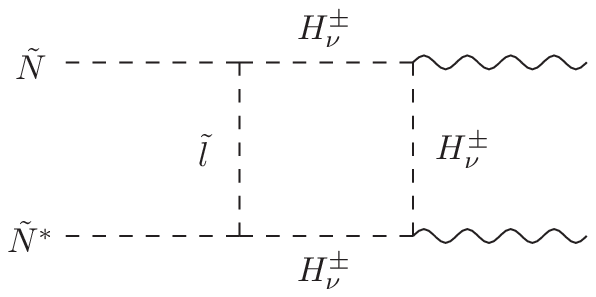}
    & & &
  \includegraphics[width=0.3\textwidth]{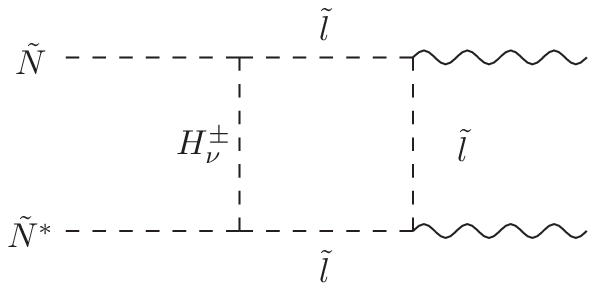}
    & & &
  \includegraphics[width=0.33\textwidth]{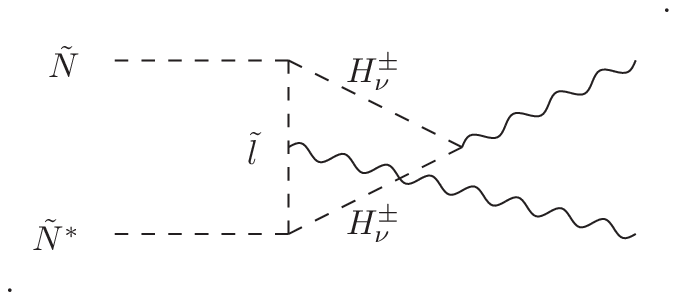}  
  \end{tabular}
  \end{center}
  \caption{Box loop diagrams for the annihilation of right-handed sneutrinos.}
  \label{fig:Box}
\end{figure}
The charged components of the $H_\nu$ scalar doublet and charged scalar fermions make the triangle or box 
 loop-diagram shown in Figs.~\ref{fig:Triangle} and~\ref{fig:Box}, and the two photons can be emitted from the internal charged particles. 
In this case, the photons have line spectrum with energy 
\dis{
E_{2\gamma} = M_{\tilde N}.
}
In the triangle diagrams, $H_\nu$ and $\tilde{l}$ are in the loop and the charged lepton or charged Higgs
 have a quartic coupling with photons. 
In this analysis we neglect those fermion loop processes because 
 the triangle diagrams with fermions are helicity suppressed and
 its leading terms are dimension-six operators.
For the box diagrams, the charged Higgs $H_\nu^\pm$ and charged sleptons in the loop emit two photons. 
Thus the amplitude is the sum of all the contributions
\dis{
{\cal M}_{2\gamma} = {\cal M}^{\text{Tringle}}_{2\gamma}+{\cal M}^{\text{Box}}_{2\gamma}.
\label{twogamma}
}
Since DM is nonrelativistic we can ignore the momentum of DM. Assuming $M_{H_{\nu}}, M_{\tilde l}\gg M_{\tilde N}$,
 we obtain the annihilation cross section to two photons via one loop as
\dis{
\VEV{\sigma v}_{2\gamma}
=\frac{|M|^2_{2\gamma} }{32\pi M^2_{\tilde N} }
 \simeq \frac{ \alpha^2_{\rm em}}{8\pi^3}  \frac{y_\nu^4 (A_\nu^2+\mu'^2)^2}{M_{\tilde l}^4} \frac{4}{M^2_{\tilde N}}.
}
 in the limit of $M_{H_\nu}=M_{H_\nu'}=M_{\tilde l}$ for simplicity.
 
The gamma-ray line spectrum can also be produced from the dark matters annihilation
 into $Z\gamma$ through box one-loop. 
The energy of the photons produced in this process is
\dis{
E_{1\gamma} \simeq M_{\tilde N}\left(1 - \frac{M_{Z}^2}{4 M^2_{\tilde N}}\right).
\label{onegamma}
}
The annihilation cross section is approximately given by
\dis{
\VEV{\sigma v}_{1\gamma} \simeq \frac{\alpha^2_{\rm em}}{8\pi^3\cos^2\theta_{\rm w}}  \frac{y_\nu^4( A_\nu^2+\mu'^2)^2}{M_{\tilde l}^4} \frac{4}{M^2_{\tilde N}} \left(1 - \frac{M_{Z}^2}{4 M^2_{\tilde N}}\right),
}
where $\theta_{\rm w}$ is Weinberg mixing angle and we used $M_{H_\nu}=M_{H_\nu'}=M_{\tilde l}$.

Recently a tentative indication of gamma-ray line using Fermi-LAT data was reported
 in~\cite{Bringmann:2012vr,Weniger:2012tx}. 
When it is interpreted in terms of dark matter particles annihilating into a photon pair,
the observation implies a dark matter mass of~\cite{Weniger:2012tx} 
\dis{
m_{\chi} = 129.8 \pm 2.4^{+7}_{-13}\, \gev,
}
and a partial annihilation cross section of 
\dis{
\VEV{\sigma v}_{\chi\chi \rightarrow \gamma\gamma}=(1.27\pm0.32^{+0.18}_{-0.28} )\times 10^{-27}\, cm^3 s^{-1}, 
}
when using the Einasto dark matter profile.

The right-handed scalar neutrino dark matter in this model can explain this observation perfectly
with mass $M_{\tilde N} =130\gev$.  
The gamma-ray line signal in the Fermi-LAT can be explained  when 
\dis{
\VEV{\sigma v}_{(\tilde{N}\tilde{N}^* \rightarrow \gamma\gamma )}= \VEV{\sigma v}_{(\chi\chi \rightarrow \gamma\gamma  )}^{\text{ Fermi-LAT}}.
}
This corresponds to the coupling
\dis{
y_\nu^4( A_\nu^2+\mu'^2)^2 \simeq  1.8 \, M_{\tilde l}^4,
}
where we used $M_{\tilde N} =130\gev$ and $\alpha_{em}=1/127$.  
For this mass of DM, we have another gamma-ray line at $E_\gamma =114 \gev$ but the flux is reduced
 by half that of two gamma line at $130 \gev$.

\subsection{Another constraint}

Here we note the constraint from direct dark matter searches.
The relevant processes for sneutrino DM direct searches are $Z$ boson and
 Higgs boson exchange.
As we note in Eq.~(\ref{LRmixing}),
 due to extremely small left-right mixing of sneutrinos,
 the couplings are considerably suppressed by ${\cal O}( (m_{\nu}/M_{\rm SUSY})^2 )$.
%
However, there is an additional $Z$ boson exchange process induced at one loop
 by the left-handed (LH) sneutrino and neutrinophilic Higgs ($H_{\nu}$ or $H_{\nu'}$)
 or those corresponding to $SU(2)$ doublet charged particles. 
In fact, the effective coupling between right-handed sneutrino $\tilde{N}$ and a quark
 through the Z boson induced by the LH sneutrino and $H_{\nu}$-like Higgs boson loop is 
\begin{equation}
 {\cal L}_{\rm{ int}} =
 \left(\frac{y A_{\nu}}{4 \pi}\right)^2 \frac{1}{12 M^2}\left(\frac{- g_2}{2 \cos\theta_W}\right)
 \frac{g_2(T-Q \sin\theta_W^2)}{M_Z^2\cos\theta_W}
 2p_{\mu} \tilde{N}\tilde{N}^* \bar{q}\gamma^{\mu} q ,
\end{equation}
 where $p$ is the four momentum of $\tilde{N}$ in the nonrelativistic limit, $M_Z$ is the Z boson mass,
 $g_2$ is the $SU(2)_L$ gauge coupling, $\theta_W$ is the Weinberg angle, and
 $M$ is the mass of the LH sneutrino and $H_{\nu}$-like Higgs boson.
For the TeV scale mass of those particles inside loops, we find the cross section with a nucleon as
\begin{equation}
 \sigma^{{\rm SI}} \simeq 10^{-9} {\rm pb} .
\end{equation}
Hence, in fact, it will be possible to explore this model
 by direct dark matter search experiments in the near future.

\section{Conclusion}
\label{conclusion}

We have shown that a Dirac right-handed sneutrino
 with neutrinophilic Higgs doublet fields is a weakly interacting massive particle and
 a viable dark matter candidate.
This is because neutrino Yukawa couplings can be as large as of the order of unity
 in models with neutrinophilic Higgs where
 the smallness of neutrino masses is explained by the small $H_{\nu}$ VEV. 
The promising signature of this sneutrino comes from the indirect detection of dark matter,
 especially gamma-ray lines. 
One-loop annihilation cross section into $\gamma\gamma$ can be larger than
 the cross section of the helicity suppressed tree-level annihilation into fermions.
Hence we can expect a large gamma-ray line signal, and 
 for instance, signals which might have been observed in the Fermi-LAT
 can be explained by its annihilation.

\section*{Acknowledgments}

We thank Seodong Shin for valueable comments.
K.-Y.C. was supported by the Basic Science Research Program through the National Research Foundation of Korea (NRF) funded by the Ministry of Education, Science and Technology Grant No. 2011-0011083.
K.-Y.C. acknowledges the Max Planck Society (MPG), the Korea Ministry of
Education, Science and Technology (MEST), Gyeongsangbuk-Do and Pohang
City for the support of the Independent Junior Research Group at the Asia Pacific
Center for Theoretical Physics (APCTP).
This work of O.S. is in part supported by
 scientific research grants from Hokkai-Gakuen.



\end{document}